\documentclass[preprintnumbers,pre,onecolumn]{revtex4}
\usepackage{amsmath}
\usepackage{graphicx,caption,subcaption,float}
\usepackage[symbol]{footmisc}
\begin{document}

\author{April Pease$^{1}$}  

\author{Korosh Mahmoodi$^{1}$\footnote{Corresponding author: free3@yahoo.com}}
\author{Bruce J. West$^{2}$}
\affiliation{1) Center for Nonlinear Science, University of North Texas, P.O. Box 311427,
Denton, Texas 76203-1427, USA\\}
\affiliation{2) Army Research Office, Research Triangle Park, NC, United States\\}

\begin{abstract}
We present a technique to search for the presence of crucial events in
music, based on the analysis of the music\ volume. Earlier work on this
issue was based on the assumption that crucial events correspond to the
change of music notes, with the interesting result that the complexity index
of the crucial events is $\mu \approx 2$, which is the same inverse
power-law index of the dynamics of the brain. The search technique analyzes
music volume and confirms the results of the earlier work, thereby
contributing to the explaination as to why the brain is sensitive to music,
through the phenomenon of complexity matching. Complexity matching has
recently been interpreted as the transfer of multifractality from one
complex network to another. For this reason we also examine the
mulifractality of music, with the observation that the
multifractal spectrum of a computer performance is significantly narrower
than the multifractal spectrum of a human performance of the same musical
score. We conjecture that although crucial events are demonstrably important
for information transmission, they alone are not sufficient to define
musicality, which is more adequately measured by the multifractality
spectrum.

\textit{Keywords\/{: Crucial events, Multifractality, Complexity matching, Musicality, 1/f noise}}

\end{abstract}

\title{Complexity Measures of Music}
\maketitle

\section{Introduction}

One of the outstanding mathematicians of the twentieth century, George
Birkhoff, argued that the aesthetics of art have mathematical, which is to
say a quantitative, measure. The structure in various art forms, music in
particular, that he discussed in his book \cite{birkhoff33} was largely
overlooked by other scientists until the last quarter of the twentieth
century, when Mandelbrot introduced the scientific community to fractals 
\cite{mandelbrot77} and his prot\'{e}g\'{e}e Voss applied these ideas to the
mathematical analysis of music. Voss and Clark \cite{voss1,voss2} used
stochastic, or \textit{1/f}, music, in which notes are selected at random
and the frequency with which a particular note is used is determined by a
prescribed distribution function, to gain insight into the structure of more
conventional music. They determined that a variety of musical forms, jazz,
blues, classical, have a blend of regularity and spontaneous change
characteristic of \textit{1/f}-music. Aesthetically pleasing music was
found to have a \textit{1/f}$^{\alpha }$ spectrum, with an inverse power-law
index in the interval $0.5<\alpha <1.5$, thereby connecting the structure of
music to the physical phenomena of \textit{1/f}-noise \cite{voss3}. 

In 1987 the newly developed concept of self-organized criticality (SOC) was
used by Bak \textit{et al}. \cite{bak87} to explain the source of \textit{1/f%
}-noise. Subsequently, \textit{1/f}-noise has been found to be a ubiquitous
property of complex networks near criticality, such as the brain. This
suggests an exciting connection with the problem of cognition \cite{anderson}
because $1/f$ noise may represe
nt the brain self-organizing through a
vertical collation of the body's spontaneous physiological events. Soma 
\textit{et al}. \cite{soma03} have shown that the brain is more sensitive to 
\textit{1/f}-fluctuations than to other forms of noise, resulting in higher
information transfer rates in the visual cortex \cite{yu05}, pain-relief
efficiency by electrical stimulation \cite{takakura79} and enhanced
efficiency by biological ventilators \cite{mutch05}. West \textit{et al}. 
\cite{west08} speculate that there is a complexity matching between Mozart's
music (\textit{1/f}-composition), the brain's organization (\textit{1/f}%
-complex network) and the hearbeat (another \textit{1/f}-process), to
explain the result of Tsuruoka \textit{et} \textit{al}. \cite{tsuruoka07}
that listening to Mozart has the effect of inducing \textit{1/f}-noise on
heart beating. This also supports the conjecture that music mirrors the mind 
\cite{bianco} in that its complexity is a reflection of the $1/f$-complexity
of brain cognition.

The observation that listening to Mozart's music enhances the reasoning
skills of students \cite{nature} contributed to the ever-expanding circle of
research interest centered on the possible complexity matching between
Mozart's music and brain function. This is a throny problem having aspects
of a number of fundamental human issues, including but not limited to
creativity, free will, determinism and randomness \cite{wang}. Our purpose
here is to present a mathematical theory that explains these interesting
aspects of music, which picks up where the above mentioned popular works
leave off. 

The approach presented herein uses the concept of a crucial event as a
fundamental building block for the underlying time series, resulting in 
\textit{1/f}-variability being the signature of complexity. The theory is an
application of the recent work of Mahmoodi \emph{et al} \cite{korosh}, which
contains an intuitive description of crucial events and develops a
generalized form of SOC, \textit{self-organized temporal criticality}
(SOTC), based on the dynamics of complex networks. Experimental observation
shows that moving from physics to biology is signaled by the emergence of
the breakdown of ergodic behavior with increasing complexity. Ergodic
behavior is one of the foundational assumptions of statistical physics, that
being that time averages of system vaiables produce results equivalent to
those obtained from ensemble averages of those variables \cite{grigolini}.
Its almost ubiqutous breakdown in complex system came as a surprise.

Ergodicity breakdown is caused by complex fluctuations being driven by
crucial events. The time interval between consecutive crucial events are
statistically independent and described by a markedly non-exponential
waiting-time probability density function (PDF). To realize the temporal
complexity of crucial events requires the concept of an intermediate
asymptotic region, characterized by an inverse power law (IPL) with index $%
\mu <3$. These events are renewal \cite{allegrini1/f} in the sense that
their occurrence invokes a total rejuvenation of the system implying that
sequential renewal events occur at times having no correlation with the
times of occurrence of preceeding events. SOTC shows that beyond the
intermediate asymptotic region an exponential time region appears that
entails the system recovering the normal ergodic condition in the long-time
limit. This exponential truncation, generated by the same cooperative
interaction responsible for the IPL nature of the intermediate asymptotic
region, is often confused with the effects produced by the finite size of
the observed time series.

We conjecture that music, being the mirror of mind, naturally reflects the
brain's dynamics, which is a generator of $1/f$ -fluctuations \cite
{allegrini1/f}, thereby confirming the early observations of Voss and Clarke 
\cite{voss1,voss2,voss3}. However, the arguments adopted by these pioneers
are based on the assumption that $1/f$-noise is generated by fluctuations
with very slow, but \emph{stationary} correlation functions. Whereas, the
crucial events emerging from the statistical analysis of the time series
generated by the brain \cite{allegrini1/f}, on the contrary, have
non-stationary correlation functions. The significance of SOTC modeling is
that the crucial events generated are the same as the $1/f$-noise produced
by the brain, that is, the fluctuations have non-stationary correlation
functions.

The importance of crucial events for music composition was recognized in two
earlier publications of our group \cite{davidadams,lestocart}; the first
paper illustrates an algorithm for composing music based on crucial events.
The present paper is closer to the main goal of the second publication,
which was the detection of crucial events in existing music composition.
Vanni and Grigolini \cite{lestocart} assummed that the time at which a note
change occurs is a crucial event and found that the IPL index was $\mu
\approx 2$. Herein we adopt a different criterion for detecting crucial
events, one based on the observation of music volume. This technique was
inspired by a method developed and used by Kello \cite{kello}, who, however,
did not evaluate the time interval between two consecutive events.
Consequently, the question of whether or not the events are crucial events
was left unanswered. 

Herein we confirm that music is driven by crucial events. Our analysis
establishes a significant difference between computer and human performance
of the same music score, which is not surprising. The computer plays the
notes as written by the composer, without interpretation. Humans, on the
other hand, bring all their knowledge, experience and feeling for the music
to their performance. The computer can provide the heart of the music, but
only a human can make the heart beat. 

We also make a preliminary attempt at establishing a connection between
crucial events and multifractality. A time series without a characteristic
time scale can be characterized by a scaling exponent, the fractal
dimension. An even more complex time series can have a time-dependent
fractal dimension, resulting in a spectrum of fractal dimensions. This
spectrum defines a multifractal time series and the width of the
multifractal spectrum is a measure of the varability\ of the time series
scaling behavior.

\section{In search of crucial events}

According to the theoretical perspective established in earlier work \cite%
{crucialevents} we define \emph{crucial events}, as events for which the
time interval between two consecutive events is described by a waiting-time
PDF $\psi (\tau )$ with the asymptotic IPL structure: 
\begin{equation}
\psi (\tau )\propto \frac{1}{\tau ^{\mu }},  \label{definition}
\end{equation}%
with an IPL index $\mu <3$. The time intervals between two different pairs
of consecutive events are not correlated 
\begin{equation}
\left\langle \tau _{i}\tau _{j}\right\rangle \propto \delta _{ij},
\end{equation}%
where the bracket indicates an average over the waiting-time PDF. The
occurrence of crucial events establishes a new kind of
fluctuation-dissipation process \cite{crucialevents2} and the transport of
information from one complex system $P$ to another complex system $S$ is
determined by the influence that the crucial events of $P$ exert on the time
occurrence of the crucial events of $S$ \cite{complexitymanagement}
(complexity management).

\begin{figure}[ptb]
\begin{center}
\includegraphics[width=0.5\linewidth]{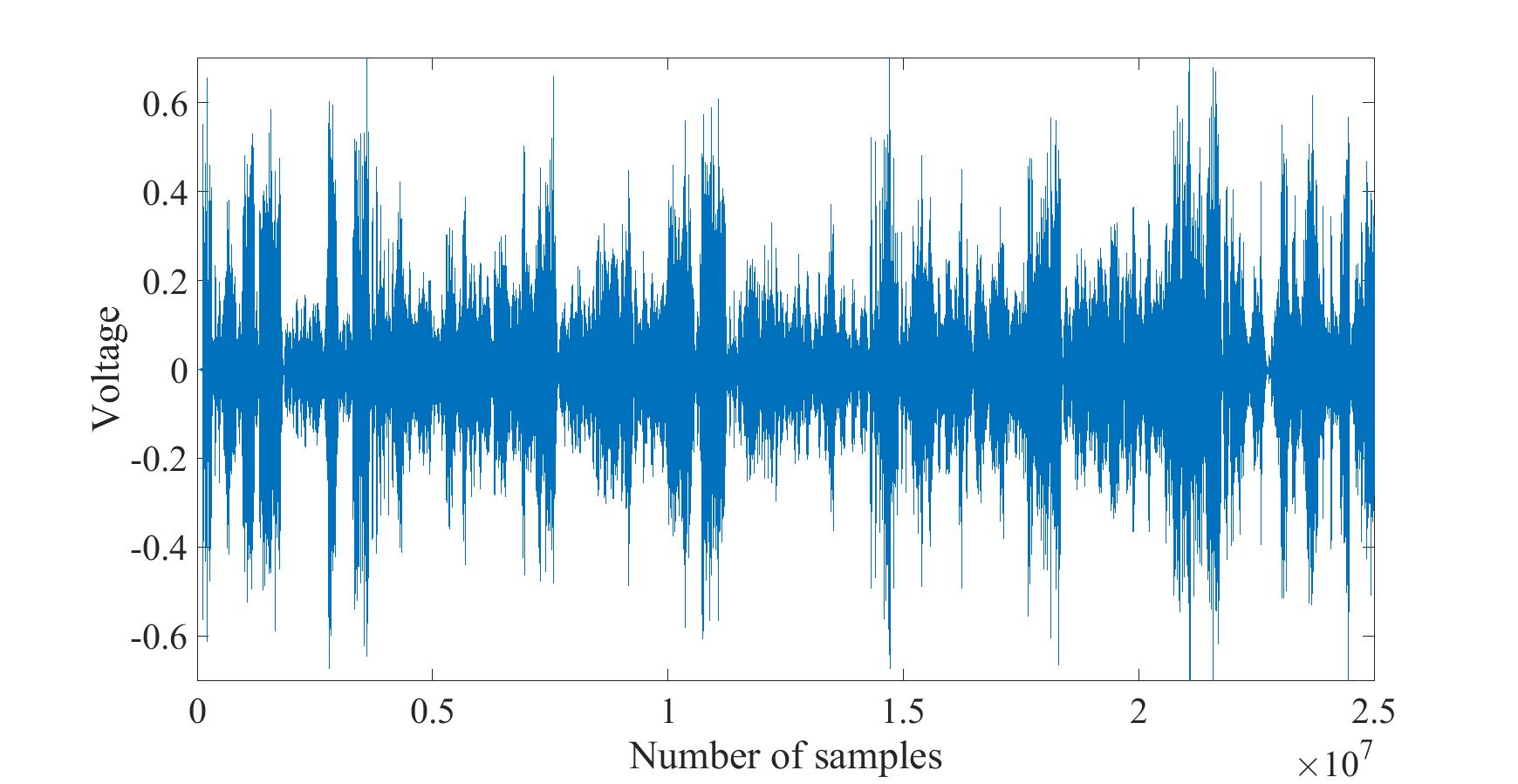}
\end{center}
\caption{An example music signal.}
\label{first}
\end{figure}

To search for renewal events in music we adopt a method that Kello used at
the 2016 Denton Workshop \cite{kello} to record music amplitude and turn it
into a sequence of events significantly departing from a homogeneous Poisson
sequence of events. We assume that the intersections of the music signal
with a suitably selected threshold line may correspond to the occurrence of
a sequence of renewal events. We apply this analysis technique to the time
series resulting from both computer and human performances of a music
composition. The computer performance consisted of programmed MIDI (Musical
Instrument Digital Interface) file and a FLAC recording (Free Lossless Audio
Codec) provided a human performance. In Fig. \ref{first} we depict the music
selection for human performance of Mozart's \textit{%
Concerto for Flute, Harp, and Orchestra, Allegro}. The music signal was
sampled at 44100 samples per second.

\begin{figure}[tbp]
\begin{center}
\includegraphics[width=0.5\linewidth]{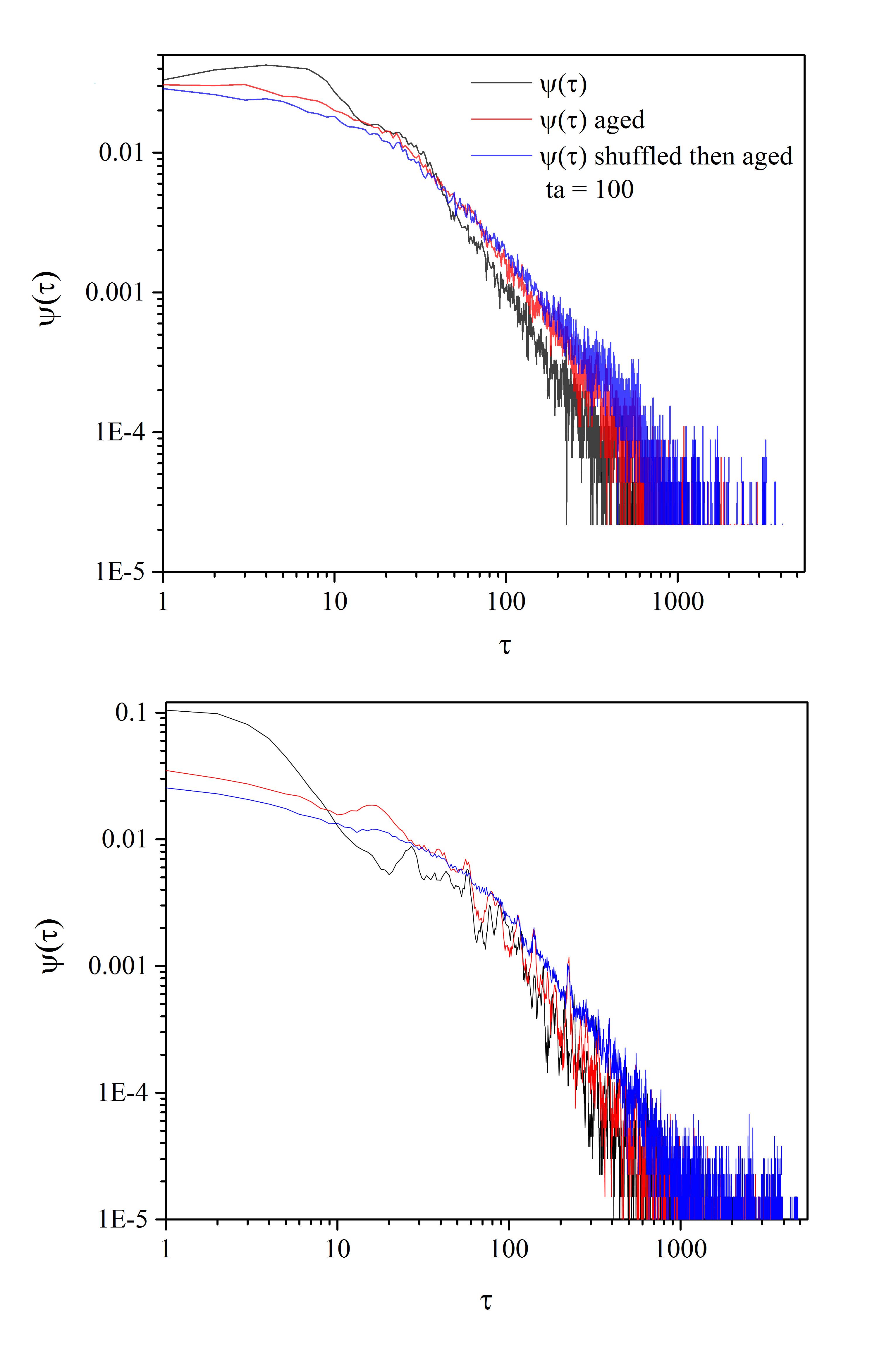}
\end{center}
\caption{The music signal waiting-time PDF is plotted vers time. The top
panel belongs to the Human performance and the bottom belongs to the
Computer performance. In each, the black line represents the waiting time
PDF of the time intervalss between two consecutive crossings of the
threshold. The red and blue lines represent both the waiting-time PDF of the
aged time series, and the shuffled then aged time series, respectively. The
size of the window used for the aging experiment is \textit{t}$_{a}$=100.}
\label{extraextra2}
\end{figure}

The IPL in Fig. \ref{extraextra2} have slopes of  $\mu $ = 2.07 and 2.2 for
the human and computer performances, respectively (shown in black). The red
and blue curves were evaluated using the aging experiment \cite%
{crucialevents}. Red is the waiting time PDF of the time series after being
aged. Blue has the $\tau $'s of the time series that are first shuffled and
then aged. As can be seen in the figure, both red and blue more or less
overlap one another. We interpret this overlap to mean the events defined by
the crossings are renewal and are therefore crucial events.

\section{Power spectrum}

Another way to establish that the events detected are renewal is to evaluate
the spectrum $S(\omega )$ to detemine if it is IPL. This is so because a
signal hosting crucial events may give the impression of being random.
Actually, that signal, as a consequence of hosting crucial events becomes a
fluctuating time series, characterized by a non-stationary correlation
function. The lack of stationarity is a consequence of ergodicity breakdown
becoming perennial when $\mu <3$. \cite{mirko}.

According to \cite{mirko}, the spectrum of fluctuations in that case cannot
be derived from the Wiener-Khintchine theorem, relying as it does, on the
stationarity assumption. It is necessary to take into account that in both
caces ($\mu =2.07$ and $\mu =2.2$), the average time interval between two
consecutive events diverges, thereby making non-stationary the process
driven by the crucial events. This anomalous condition leads to a spectrum
that is dependent on the length of the time series $L$ \cite{mirko}: 

\begin{equation}
S(\omega )\propto \frac{1}{L^{2-\mu }}\frac{1}{\omega ^{\beta }},
\label{inaccuracy}
\end{equation}%
with the IPL index 
\begin{subequations}
\begin{equation}
\beta =3-\mu .  \label{beta1}
\end{equation}%

This result was obtained by going beyond the Wiener-Khintchine theorem
adopted by Voss and Clarke in their analysis, but which cannot be applied to
our condition if we make the reasonable assumption, based \ on the results
depicted in Fig. \ref{extraextra2},  that the events detected using the
adaptation of Kello's method, are renewal. If they are renewal and they
drive the signal $\xi (t)$, namely the music intensity, then the spectrum $%
S(\omega )$ is expected to follow the prescription of Eq. (\ref{inaccuracy}%
). In the case where the process yields a slow, but stationary correlation
function, we would have $\beta <1$ \cite{mirko}. Evaluating the power
spectrum in this case becomes computationally challenging because, as shown
by Eq. (\ref{inaccuracy}), the noise intensity decreases with increasing $L$%
, the length of the time series. Nevertheless, the results depicted in Fig. %
\ref{humanSPECTRUM} yield the IPL index $\beta \approx 1$, and Eq.(\ref%
{beta1}) yields $\beta = 3-2=1,$ and the agreement between the results
is very encouraging.

\begin{figure}[tbp]
\begin{center}
\includegraphics[width=0.5\linewidth]{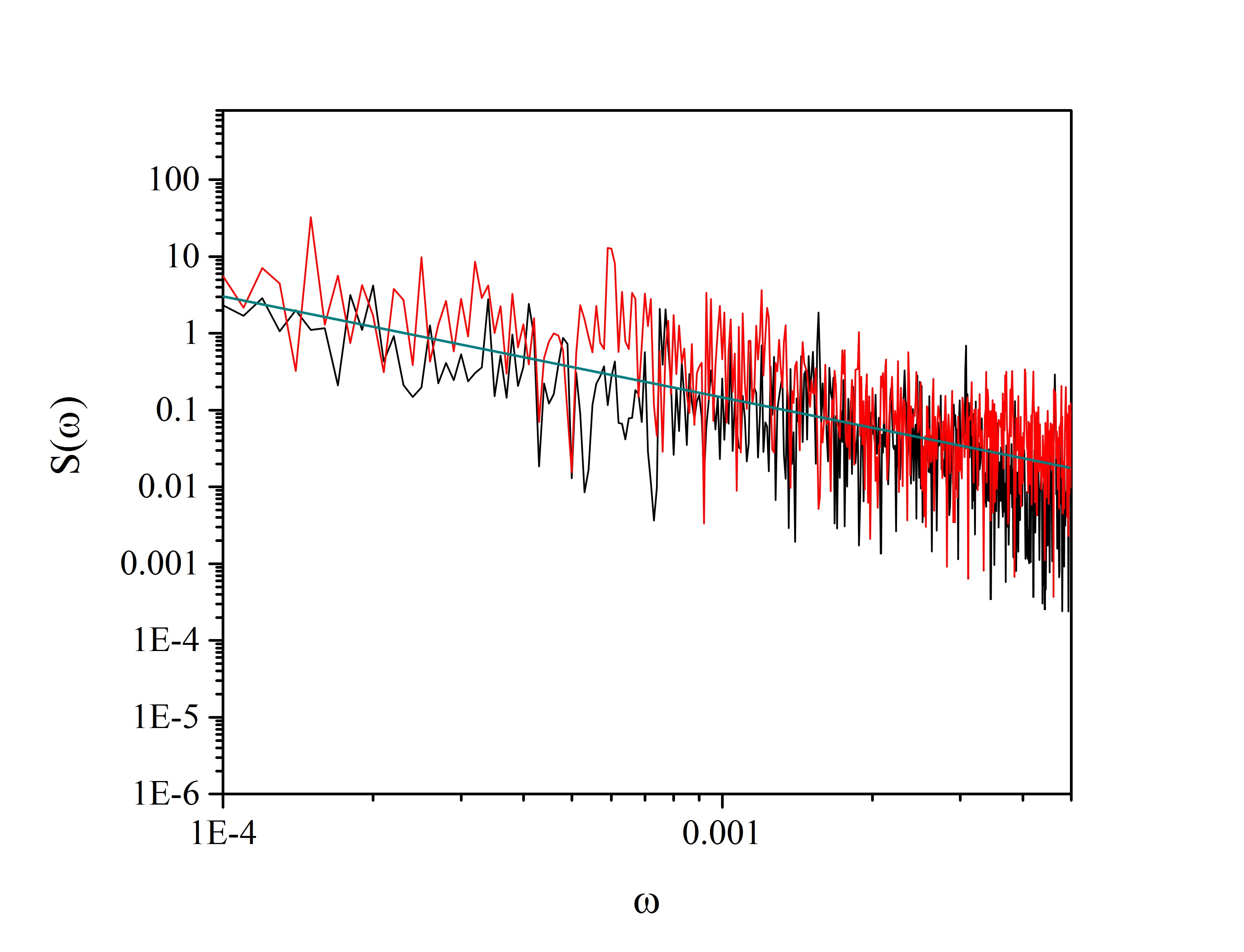}
\end{center}
\caption{ Power spectra for humman (black) and computer (red) data. The IPL
indices are approximately one.}
\label{humanSPECTRUM}
\end{figure}

Both results satisfactorily support the claim that the events revealed using
the threshold method are crucial events. To clarify this point Fig. \ref%
{Exponential} illustrates the waiting-time PDF of the intervals between two
consecutive crossings of the threshold line, when the threshold is set equal
to 0.002. This threshold is not large enough to filter out the events that
are not crucial. We see that these non-crucial events produce a well
pronounced exponential shoulder in the waiting-time PDF. The results of Fig. %
\ref{extraextra2} have been obtained by filtering out these non-crucial
events. We therefore conclude that the crucial events, which are the
mechanism for information transport \cite{complexitymanagement}, also have
the significant effect of determining the behavior of the spectrum for $%
\omega \rightarrow 0$.

\begin{figure}[tbp]
\begin{center}
\includegraphics[width=0.5\linewidth]{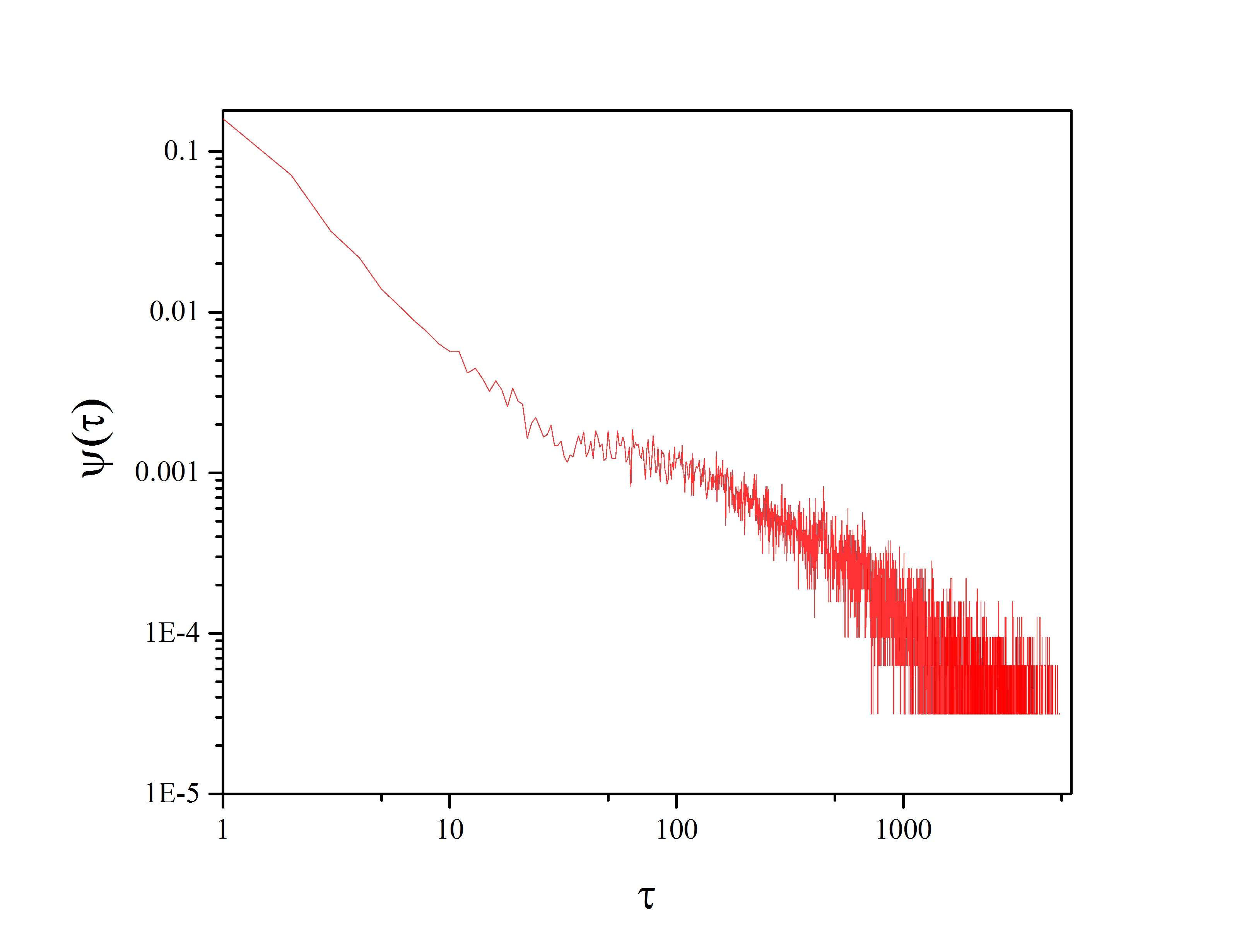}
\end{center}
\caption{ The waiting-time PDF of the human performancetime series, obtained
using a threshold of 0.002.}
\label{Exponential}
\end{figure}

\section{Multifractality}

\label{multifractality}

The discovery of $1/f$ noise in music by Voss and Clarke \cite{voss1} was
interpreted assuming the music time series is stationary, which is
consistent with Fractional Brownian Motion (FBM), and yields a mono-fractal 
\cite{mandelbrot77}. However, the present work goes beyond \cite{voss1} and
the ergodic assumption, by taking a non-stationary approach consistent with
multifractality. 

Using the method of Multifractal Detrended Fluctuation Analysis (MF-DFA) 
\cite{Kantelhardt} to analyze each time series of the two performances gives
the results shown in the Fig. \ref{multimusic}.  The computer performance
yields the narrow multifractal distribution, whereas the multifractal
distribution of the human performance is significantly  broader. This
notable difference between the multifractal spectra indicates different
levels of complexity in the two performances. The narrowness of the computer
performance suggests a strict adherence to the single fractal dimension and
consequenty less complexity then in the human performance. In fact, the
difference between the two performances may be better described by what the
computer performance lacks compared to the human performance. This extra
information, or musicality, contained in the human performance includes
specific techniques that add to the complexity of the music through subtle
variation in timing, intonation, articulation, dynamics, etc., which are
likely a better match to the brain's complexity. The human performance is
largely more aestheticly pleasing to the listener than is the computer
performance.

Suppose the brain of Mozart contains a certain complexity, which is well
described by SOTC as a generator of $1/f$-noise. Then Mozart transcribes his
complexity, albeit incompletely, into the music score of the chosen
selection. To recover this lost musicality, the human performer injects
their own interpretation of the lost complexity using their specific
performance techniques. Conversely, the computer performance is unable to
interpret this lost component and delivers exactly what was transcribed,
resulting in less variability in complexity. The computer performance is a
record of the brain of Mozart even if Mozart himself would have produced a
broader multifractal distribution when the piece was performed.

\begin{figure}[ptb]
\begin{center}
\includegraphics[width=0.5\linewidth]{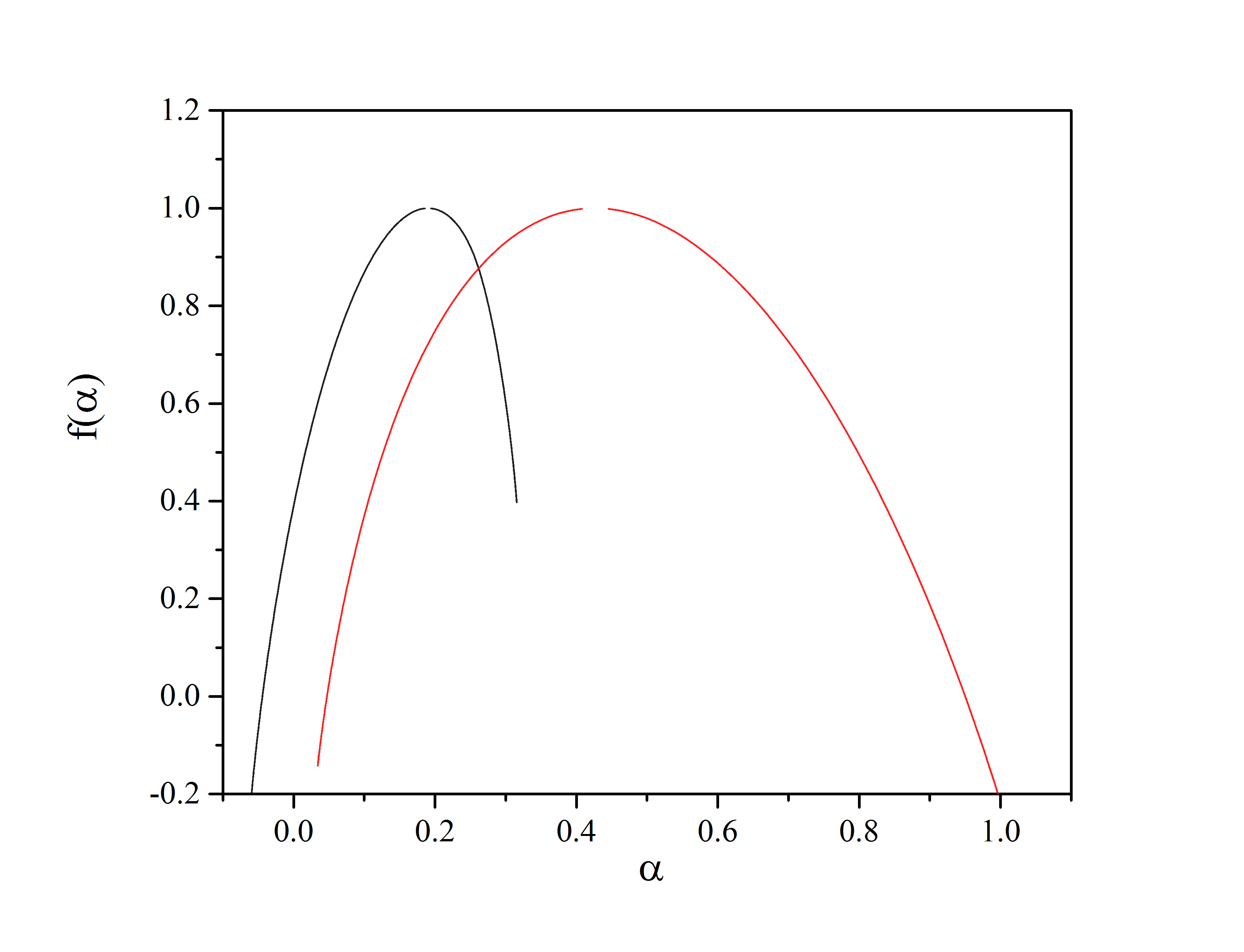}
\end{center}
\caption{The black line represents multifractal spectrum of the computer
performance, and the red represents the human performance's multifractal
spectrum.}
\label{multimusic}
\end{figure}

\section{Concluding Remarks}

\label{Conclusion}

Crucial events exist in the changing of notes in music, as found by Vanni 
\emph{et al} \cite{lestocart}. Similarly, this analysis done differently by
analyzing the music signal, the change in volume, leads to the same
conclusion (through the analysis of a different aspect of the music). This
difference in analysis is very important because the statistical analysis of
the dynamics of the brain \cite{allegrini1/f} shows that the brain is a
generator of crucial events with the same IPL index. Additionally, we found
that there is a noticeable difference in the fractal measures between human
and computer performances.

Music is aesthetically pleasing to the brain \cite{davidadams,lestocart}
because of the crucial events described by $\mu $ = 2. Multifractality may
provided a clearer picture of which performance of Mozart was more pleasing.
The difference in musicality is obvious to the listener's ear and this
difference can be quantified through the narrower and broader multifractal
spectra. The increasing aesthetics of music favors a broader multifractal
spectrum. Indeed, multifractality describes an additional measure of
complexity. Crucial events measure complexity of the intermediate
asymptotics, whereas multifractality contains additional information, beyond
the intermediate asymptotics, regarding the transient region and exponential
truncation. All this subtlety in composition is experienced by the brain
through the transfer of the music time series' multifracality.

\textbf{\emph{Acknowledgments}}. AP and KM warmly thank ARO and Welch for
support through Grant No. W911NF-15-1-0245 and Grant No. B-1577,
respectively.
\end{subequations}



\end{document}